\documentclass[useAMS,usenatbib]{mn2e}
\usepackage{graphicx}

\title[Spectral states of IGR J17480--2446]{Terzan 5 transient IGR J17480--2446:
variation of burst and spectral properties with spectral states}
\author[Chakraborty, Bhattacharyya and Mukherjee]{Manoneeta Chakraborty$^{1}$\thanks{E-mail: manoneeta@tifr.res.in}, Sudip Bhattacharyya$^{1}$\thanks{E-mail:
sudip@tifr.res.in} and Arunava Mukherjee$^{1}$\thanks{E-mail:
arunava@tifr.res.in} \\
$^{1}$Department of Astronomy and Astrophysics, Tata Institute
of Fundamental Research, Mumbai 400005, India}

\begin{document}

\date{
}

\pagerange{\pageref{firstpage}--\pageref{lastpage}} \pubyear{2010}

\maketitle

\label{firstpage}
\begin{abstract}
We study the spectral state evolution of the Terzan 5 transient neutron star
low-mass X-ray binary IGR J17480--2446,
and how the best-fit spectral parameters and burst properties evolved with these states,
using the {\it Rossi X-ray Timing Explorer} data. As reported by other authors,
this is the second source which showed transitions between atoll state and `Z' state.
We find large scale hysteresis in the almost `C'-like hardness-intensity track of the source in the 
atoll state. This discovery is likely to provide a missing piece
of the jigsaw puzzle involving various types of hardness-intensity tracks from `q'-shaped for
Aquila X-1, 4U 1608--52, and many black holes to `C'-shaped for many atoll sources.
Furthermore, the regular pulsations, a diagonal transition between soft and hard states,
and the large scale hysteresis observed from IGR J17480--2446 
argue against some of the previous suggestions involving magnetic field about
atolls and millisecond pulsars. Our results also suggest that 
the nature of spectral evolution throughout an outburst does not, at least entirely,
depend on the peak luminosity of the outburst.
Besides, the source took at least a month to trace the softer banana state,
as opposed to a few hours to a day, which is typical for an atoll source. 
In addition, while the soft colour
usually increases with intensity in the softer portion of an
atoll source, IGR J17480--2446 showed an opposite behaviour.
From the detailed spectral fitting we conclude that a blackbody+powerlaw model
is the simplest one, which describes the source continuum spectra well
throughout the outburst. We find that these two spectral components were plausibly 
connected with each other, and they worked together to cause the
source state evolution. Spectral parameters smoothly changed as IGR J17480--2446
transitioned between atoll state and `Z' state, and thermonuclear bursts disappeared
in the softer parts of `Z' tracks. Finally, based on the burst properties, 
we suggest that IGR J17480--2446 is somewhat an analogue 
of the only other clock-like burster GS 1826--238, 
and the former showed much smaller burst intervals plausibly because it is a pulsar.
\end{abstract}


\section{Introduction}\label{Introduction}

The transient source IGR J17480--2446 was discovered at the onset of its outburst
during the Galactic bulge monitoring with {\it INTEGRAL} on Oct 10, 2010 \citep{Bordasetal2010}.
The source location was consistent with that of the globular cluster
Terzan 5, which motivated the identification of this source as the
known Terzan 5 transient 
and type-I (thermonuclear) burster EXO 1745--248 \citep{Bordasetal2010, LinaresAltamirano2010}.
Detection of thermonuclear bursts from IGR J17480--2446, with {\it INTEGRAL} on Oct 11, 2010
\citep{Chenevezetal2010} and with {\it Rossi X-ray Timing Explorer} ({\it RXTE}) on
Oct 13, 2010 \citep{StrohmayerMarkwardt2010}, supported this identification, 
and established that this source is a neutron star low-mass X-ray binary (LMXB).
However, {\it Swift} and {\it Chandra} observations revealed that this transient neutron
star LMXB of Terzan 5 is not EXO 1745--248 \citep{Heinkeetal2010, Pooleyetal2010}. 
Subsequently, \citet{Ferrignoetal2010}
suggested IGR J17480--2446 as the name of this source in recognition of its
discovery with {\it INTEGRAL}.

IGR J17480--2446 is a very interesting and unique source for the following reasons.\\
(1) Since it is a globular cluster source, its distance is relatively accurately
known ($\approx 5.5\pm0.9$ kpc; \citet{Ortolanietal2007}). Therefore, it can be a promising source
to estimate the neutron star radius from the continuum spectroscopy of thermonuclear 
bursts (see \citet{Bhattacharyya2010} and references therein). \\
(2) IGR J17480--2446 is a $\approx 11$ Hz pulsar \citep{StrohmayerMarkwardt2010}. Such a
slowly spinning neutron star is unusual for LMXBs, because we expect accretion-induced
spin-up of such stars \citep{BhattacharyavandenHeuvel1991}. This low spin rate might 
indicate that enough matter has not yet been accreted, which might have implications for 
binary evolution, or there is a spin-down mechanism, such as gravitational radiation
\citep{Chakrabartyetal2003}, or the neutron star is already in spin equilibrium
due to an unusually strong stellar magnetic field (which is unlikely considering the 
magnetic field quoted for this source; see \citet{Cavecchietal2011}). \\
(3) The source showed burst oscillations \citep{Altamiranoetal2010a}.
This is the most slowly spinning burst oscillation source,
indicating that rapid rotation is not necessary for the oscillation mechanism.
IGR J17480--2446 is also one of the very few neutron star LMXBs showing
both pulsations and burst oscillations. Consequently, this source can be 
very useful to probe these two phenomena more effectively through their
mutual interaction, as both of them originate from the stellar surface, and
can occur simultaneously (see \citet{Bhattacharyya2010} and references therein). \\
(4) Based on the correlated spectral and timing behaviour, neutron star LMXBs
can be divided into two classes: atoll and Z (see \citet{vanderKlis2006} and references therein).
IGR J17480--2446 is only the second source (after XTE J1701--462; \citet{Homanetal2010}) 
which changed from an atoll state into a Z state \citep{Altamiranoetal2010b}, 
and hence is very important to understand the evolution of neutron star LMXBs. \\
(5) After the Oct 13 burst mentioned above, IGR J17480--2446 displayed recurrent
bursts at short intervals \citep{Altamiranoetal2010b, Papittoetal2011}.
As the source intensity increased, the bursts gradually became more frequent, and eventually
disappeared, while millihertz (mHz) quasi-periodic oscillations (QPOs) appeared
\citep{Altamiranoetal2010b, Linaresetal2010}. This made IGR J17480--2446 one
of very few mHz QPO sources. The appearance of mHz QPOs supports the thermonuclear
origin of the frequent bursts, as a popular mHz QPO model interprets this timing 
feature as a signature of marginally stable nuclear burning \citep{Hegeretal2007}.
As the source intensity decayed, the bursts started becoming less frequent.
The repetition of bursts at short intervals, 
as well as the lack of significant cooling in the decay portions
of many of them, motivated \citet{GallowayZand2010} to suggest that these were
type-II bursts, i.e., they were powered by the gravitational potential energy.
However, recently \citet{ChakrabortyBhattacharyya2011}, \citet{Mottaetal2011} have performed detailed
data analysis, and argued that these bursts were likely of thermonuclear origin.
If this is true, then the unique clock-like systematic behaviour of the bursts 
from IGR J17480--2446
will be extremely important to probe the nuclear burning on neutron stars. \\
(6) Accretion disk wind has been detected from IGR J17480--2446, which
is plausibly the clearest such detection from a neutron star system \citep{Milleretal2011}.
Note that this wind is more commonly observed from the stellar-mass black hole systems. \\
(7) \citet{Bhattacharyya2010} discussed ten methods to measure the parameters
of neutron stars in LMXBs, and emphasized that the joint application of as many
of these complementary methods as possible is required to reliably measure the
neutron star parameters.
IGR J17480--2446 showed thermonuclear bursts, all three high-frequency
timing features (pulsations, burst oscillations, kilohertz (kHz) QPOs; 
\citet{Altamiranoetal2010b}), mHz QPOs, broad spectral iron line 
\citep{ChakrabortyBhattacharyya2011, Milleretal2011} and quiescent emission
\citep{DegenaarWijnands2011}. Therefore, seven out of the ten methods
are, in principle, available for this source.

In this paper, we study the spectral state evolution of IGR J17480--2446,
and how burst properties and best-fit spectral parameters varied with the state
evolution. Here is the motivation for this study. Apart from the near-peak-intensity time, 
the source was likely to be in atoll state. An atoll source usually traces a 
`C'-like curve in the colour-colour
diagram (CD; hard colour versus soft colour)
and in the hardness-intensity diagram (HID; hard colour versus intensity;
see \S~\ref{DataAnalysisandResults}; \citet{vanderKlis2006}).
Typically, the higher hard colour extreme island state is traced out in days to weeks.
The lower hard colour banana-like portion (banana state) of the `C' track can be divided into
upper banana, lower banana and lower left banana based on
spectral and timing properties. The banana state is traced out on time scales of hours to a day
without any hysteresis \citep{vanderKlis2006}.
An atoll source moves from extreme island state to banana state via an island state,
and vice versa. \citet{Gladstoneetal2007} showed that this transition in CD occurs 
vertically for some sources (verticals), and diagonally for other sources (diagonals).
As opposed to most of the neutron star LMXBs, a transient black hole LMXB usually traces a `q'-like
large scale hysteresis curve in the HID \citep{vanderKlis2006, Belloni2009}.
This difference between the neutron star and black hole sources 
makes it difficult to relate the accretion processes in neutron star systems
and black hole systems. However, the transient atoll source Aquila X-1 and 4U 1608--52 show
`q'-like large scale hysteresis HID curves (\citet{MaitraBailyn2004, Reigetal2004, Gladstoneetal2007}; 
see also \citet{Bellonietal2007} for 4U 1636--53 tracks).
Moreover, recently \citet{MukherjeeBhattacharyya2011} has analyzed the
{\it RXTE} data from the transient atoll EXO 1745--248 of Terzan 5, and 
reported that this source traced a large scale hysteresis HID curve, which was 
intermediate between `q'-like tracks and `C'-like tracks, and hence could be 
useful to understand the spectral state evolution of both neutron star systems
and black hole systems. 
This motivated us to study the
CD/HID tracks of IGR J17480--2446, which is also a Terzan 5 transient atoll source.
We also analyze the spectra in detail in order to identify the
spectral parameters causing the observed CD/HID tracks of IGR J17480--2446.
Finally, we probe the evolution of burst properties, which will be useful
to understand these unique bursts.

\section{Data Analysis and Results}\label{DataAnalysisandResults}

The neutron star LMXB IGR J17480--2446 was observed many times (almost everyday 
between Oct 13, 2010 and Nov 19, 2010) with
{\it RXTE} during its outburst. The total observation time of this 
transient was $\approx 297$ ks 
(proposal no. 95437; 46 obsIds: 95437-01-01-00 to 95437-01-14-00).
In this section, we discuss the CD, HID, spectral properties and bursts of the source.

\subsection{Colour-colour and Hardness-intensity Diagrams}\label{CDHID}

\begin{figure*}
\centering
\includegraphics*[width=18cm]{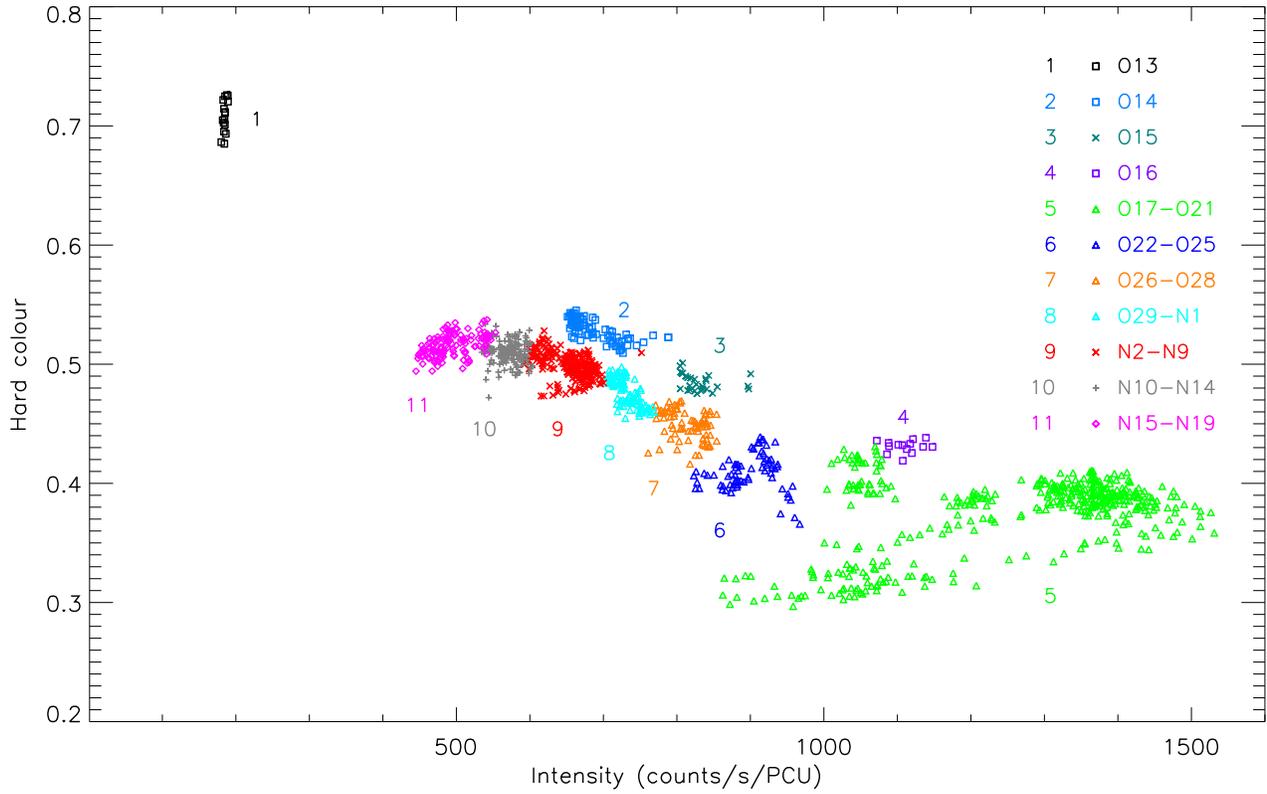}
\caption{Hardness-intensity diagram of IGR J17480--2446 using the
{\it RXTE} PCA data. Hard colour and intensity (for PCU 2) are defined
in \S~\ref{DataAnalysisandResults}.
Various temporal segments (Table~\ref{tableburst}
and \S~\ref{DataAnalysisandResults}) are shown with different
symbols and segment numbers (see Table~\ref{tableburst}). 
A list of segment numbers, symbols and the corresponding dates are also
given. In the list, `O' represents `October' and `N' represents `November'.
This figure clearly shows a large scale hysteresis in the spectral states.
\label{HID}}
\end{figure*}

We have defined hard colour and soft colour as the ratio of the background-subtracted detector counts in the 
$(9.2-18.9)/(5.7-9.2)$ and $(3.9-5.7)/(2.6-3.9)$ keV energy bands, respectively.
The intensity in HID is the background-subtracted Proportional Counter Unit (PCU) 2
count rate in the $2.6-18.9$ keV band. The CD and HID have been produced
using the entire standard-2 mode data from the top layers of PCU-2, after filtering out
the portions of bursts and data gaps.
We have been able to divide the data into 11 temporal segments (see Table~\ref{tableburst} for 
time ranges). Each segment traces a distinguishable portion of the HID track, which helps us to clearly
follow the source movement in the HID (Fig.~\ref{HID}). This figure shows that IGR J17480--2446
was in a low intensity high hard colour state on Oct 13 ($\approx 185$ counts/s/PCU and $\approx 0.70$; 
plausibly extreme island state or island state; see \S~\ref{Introduction}). 
It moved to significantly higher intensity and lower hard colour values on 
Oct 14 ($\approx 690$ counts/s/PCU and $\approx 0.53$; Table~\ref{tableburst}), 
and subsequently the intensity gradually increased and the hard colour value slowly
decreased (Fig.~\ref{HID} and Table~\ref{tableburst}). We have combined all the data during
Oct 17--21 in one segment (Fig.~\ref{HID}), because the bursts could not be clearly separated 
from the non-burst portions, and hence were not filtered out for these days. 
It is likely that, at least during a part of
this segment 5, the source was in `Z' state (\citet{Altamiranoetal2010b}; see also later).
After segment 5, the source intensity gradualy decreased from $\approx 898$ counts/s/PCU in 
segment 6 to $\approx 494$ counts/s/PCU in segment 11. At the same time, the hard colour value increased
from $\approx 0.41$ to $\approx 0.52$ (Table~\ref{tableburst}).
It is likely that during the segments 2--4 and 6--11, IGR J17480--2446 was in banana state 
(see \S~\ref{Introduction}). In these segments the source displayed a clear large scale hysteresis,
as the hard-to-soft track made of 2--4 segments is significantly above the soft-to-hard
track made of 6--11 segments in HID. In CD, however, the source did not show a 
clear hysteresis (Fig.~\ref{CD}). The hard colour and soft colour values roughly decreased during the rise of the
source outburst, and these parameters increased during the decay of the outburst.

The segment 5, being the plausible `Z' state \citep{Altamiranoetal2010b}, 
required a further investigation. 
We have produced CD and HID tracks for each day during Oct 17--21, and named
them 5a, 5b, 5c, 5d and 5e, respectively (Table~\ref{tableburst}).
In the HID, segments 5a--5d show portions of `Z'-like tracks (Figs.~\ref{CDHID1} and 
\ref{CDHID2}), which confirmed the report of \citet{Altamiranoetal2010b}. 
Such secular motions of the `Z' tracks are usual for Z sources \citep{vanderKlis2006}.
Segment 5e might be a part of a `Z' track, or it might be a part of 
banana state. Each of 5b--5d has a harder portion (5b1, 5c1, 5d1) and a softer portion 
(5b2, 5c2, 5d2; Table~\ref{tableburst}; Fig.~\ref{CDHID2}).
While for 5b--5d, IGR J17480--2446 plausibly moved from the harder to the softer portion,
only for 5b there is evidence that the source came back to the harder portion 
(Table~\ref{tableburst}). The average intensity increased from 5a to 5b, 
and then decreased gradually up to 5e. 

\subsection{Spectral Properties}\label{SpectralProperties}

We have fitted the deadtime corrected non-burst (persistent) PCU-2 top layer 
spectra in $3-15$ keV, using backgrounds produced from
the {\it RXTE} data analysis tool {\tt PCABACKEST}. We have fixed the 
neutral hydrogen column density
$N_{\rm H}$ at $3.8\times10^{22}$ cm$^{-2}$ \citep{GallowayZand2010, Kuulkersetal2003}.
We have tried two thermal models, blackbody ({\tt bbodyrad}) and multi-colour
blackbody ({\tt diskbb}), and a non-thermal model, power-law ({\tt powerlaw}),
to fit the continuum spectra. No single component model gives an acceptable fit.
Among the three two-component models (involving the above three model
components), {\tt bbodyrad+powerlaw} gives by far the best fit. Addition of a
Gaussian emission line ({\tt Gaussian}) improves the fitting significantly.
Therefore, we conclude that {\tt phabs*(bbodyrad+powerlaw+Gaussian)}
is the simplest model which provides acceptable fits throughout the outburst. The {\tt Gaussian}
represents the broad Fe K$\alpha$ line (FWHM $\sim 1.02\pm0.25$ keV, 
equivalent width $\sim 187\pm$96 eV) which plausibly originated from the inner accretion disc
(\citet{BhattacharyyaStrohmayer2007}; \citet{Bhattacharyya2010} and references therein).
However, in this paper we will discuss only the continuum spectral properties of 
IGR J17480--2446, since {\it RXTE} is not very suitable to study spectral lines.
We have fitted one spectrum for each day or each segment (whichever is smaller)
using a standard-2 mode file (time ranges are given in Table~\ref{tablespec2}). 
The corresponding best-fit parameters, unabsorbed total fluxes and ratios of unabsorbed
blackbody flux to unabsorbed powerlaw flux are given in Table~\ref{tablespec2}.
We have used this table to estimate the average values of total flux, flux ratio,
blackbody temperature and powerlaw index for each segment, and used these average values
and the hard colour and soft colour numbers given in Table~\ref{tableburst} to produce the Fig.~\ref{spectime}.
This figure shows the correlations among the evolutions of these parameters (see
\S~\ref{Discussion}).

\subsection{Bursts}\label{Bursts}

In Table~\ref{tableburst}, we have listed the burst peak flux, interval
(recurrence time), duration and fluence with the source intensity, hard colour and soft colour for each segment. 
A burst peak flux is calculated
by fitting the PCU-2 standard-2 top layer spectra, using the pre-burst emission as the
background \citep{ChakrabortyBhattacharyya2011}. The fluence of a burst is 
estimated by adding the fluxes of all adjacent time ranges in which the burst is divided.
For each day, bursts of one standard-2 file (time ranges given in Table~\ref{tablespec2}) 
has been used to estimate the average
peak flux and fluence. If a segment has more than one day (e.g., 6, 7, 8, 9, 10, 11), then
the final peak flux and fluence of the segment have been estimated by averaging over all
the days. If more than one segment exists in one day, the peak flux and fluence of such a segment
have been estimated using the appropriate standard-2 file. 
The burst intervals and durations
have been estimated with the same averaging procedure, but using the event mode
files instead of the standard-2 files. For the segment 5, the bursts could not be clearly 
identified (\S~\ref{CDHID}), and hence the peak fluxes, fluences and durations 
could not be estimated. 
However, in some cases the burst intervals could be estimated from mHz QPO frequencies
(see \S~\ref{Introduction}). These frequencies could be measured from the Leahy
power spectra produced from the event files using the standard techniques
\citep{vanderKlis1989}.

\section{Discussion and Conclusions}\label{Discussion}

\begin{figure*}
\centering
\includegraphics*[width=18cm]{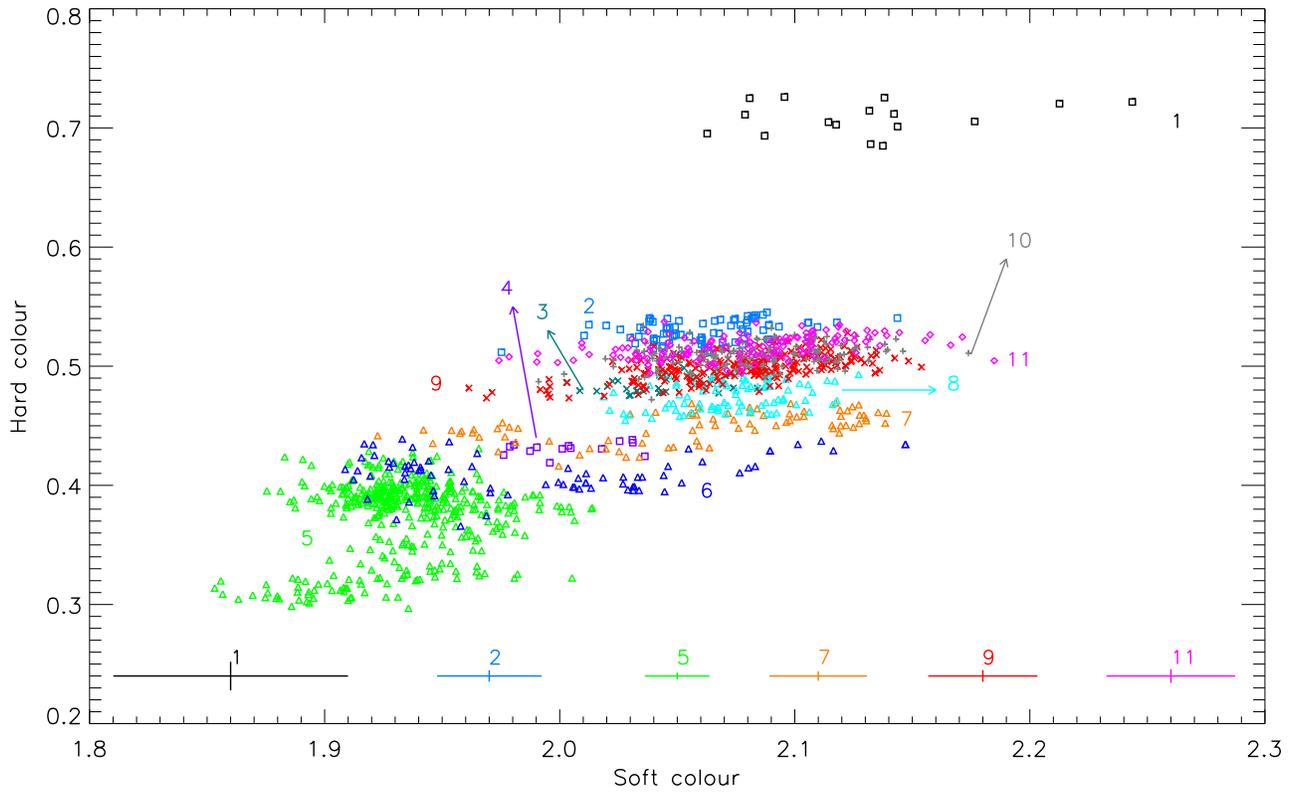}
\caption{Colour-colour diagram of IGR J17480--2446 using the
{\it RXTE} PCA data. Hard colour and soft colour are defined
in \S~\ref{DataAnalysisandResults}.
Various temporal segments (see Table~\ref{tableburst} 
and \S~\ref{DataAnalysisandResults}) are shown with different
symbols and segment numbers (see Table~\ref{tableburst}). 
Symbols are same as in Fig.~\ref{HID}. Typical $1\sigma$ error
bars for some of the segments are shown.
\label{CD}}
\end{figure*}

\begin{figure*}
\centering
\includegraphics*[width=18cm]{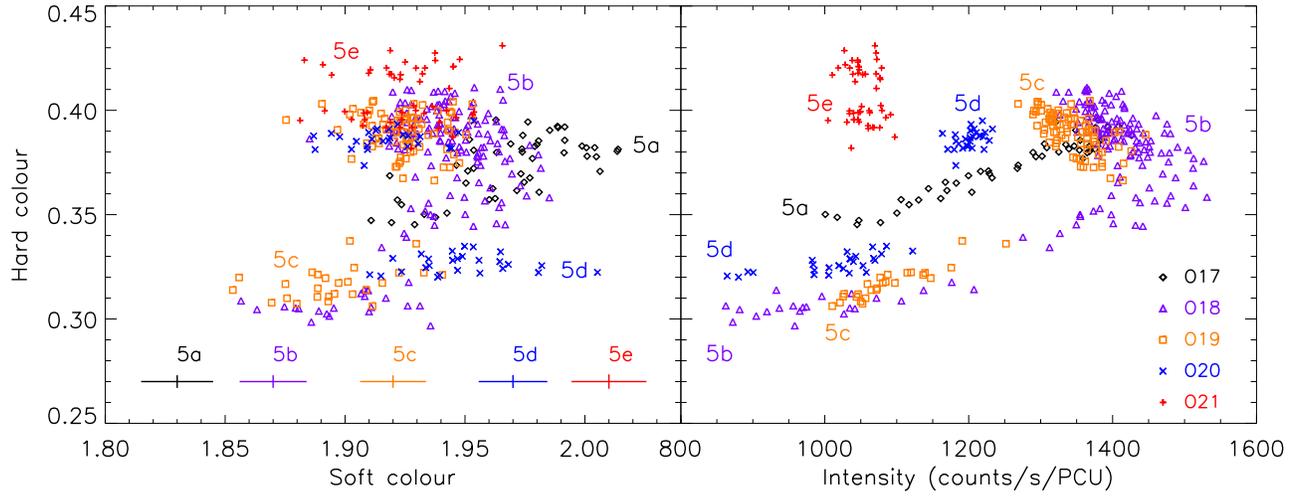}
\caption{Zoomed in colour-colour diagram (CD; left panel) and hardness-intensity diagram
(HID; right panel) of the segment 5 of IGR J17480--2446 (see Table~\ref{tableburst}
and \S~\ref{DataAnalysisandResults}). This figure suggests secular motions of `Z' track,
as the track of a given day (`O' represents `October') looks like a part of a `Z'.
Typical $1\sigma$ error bars are shown in the left panel.
\label{CDHID1}}
\end{figure*}

\begin{figure*}
\centering
\includegraphics*[width=18cm]{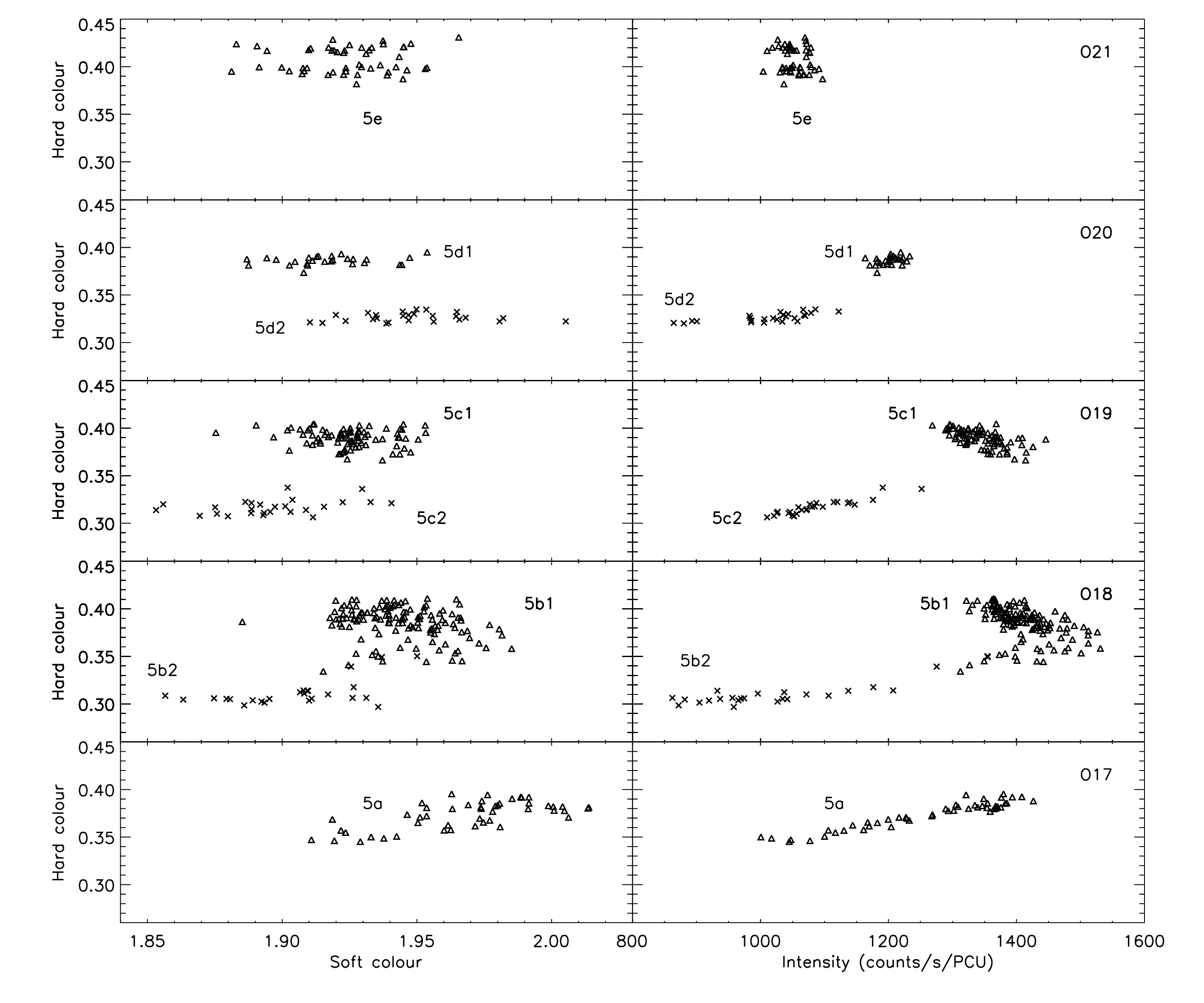}
\caption{Similar to Fig.~\ref{CDHID1}, but tracks of various days are shown in different
panels for clarity. Harder and softer parts of 5b--5d are displayed with different symbols
(see Table~\ref{tableburst}).
\label{CDHID2}}
\end{figure*}

The CD and HID tracks of IGR J17480--2446 show the uniqueness of the source. 
It took at least a month to trace the
entire banana state, while for a typical atoll source this time is hours to a day
(\S~\ref{Introduction}). While the soft colour
usually increases with intensity in the softer portion (i.e., banana state) of an 
atoll source \citep{vanderKlis2006}, IGR J17480--2446 showed an opposite
behaviour (Figs.~\ref{HID} and \ref{CD}). At the lowest soft colour and hard colour values and the highest intensities,
`Z'-like tracks in CD/HID appeared, which cleanly shows the transitions
of the source between atoll state and `Z' state.
Note that, as expected, the CD/HID `Z' tracks were the most prominent
when the average source intensity was the highest (Fig.~\ref{CDHID2}).

In the atoll state, IGR J17480--2446 showed a clear large scale hysteresis in the HID (Fig.~\ref{HID}
and \S~\ref{DataAnalysisandResults}).
This HID track with hysteresis is even more `C'-like than that of EXO 1745--248 
(\S~\ref{Introduction}). Therefore, IGR J17480--2446 could be useful to probe the 
physical differences between `q'-shaped and `C'-shaped tracks. Fig.~\ref{schematic}
shows a continuation from the `q'-shaped tracks to `C'-like tracks, i.e., how
for Aquila X-1, EXO 1745--248 and IGR J17480--2446 the large scale hysteresis can be
`q'-shaped, intermediate and `C'-like respectively. Therefore, our discovery 
of a large scale hysteresis from IGR J17480--2446 would provide a missing piece
of the jigsaw puzzle involving various types of HID tracks from `q'-shaped for
Aquila X-1, 4U 1608--52, and many black holes to `C'-shaped for many atoll sources.
IGR J17480--2446 is further important, because such sources, which show hysteresis 
like many black holes could be very useful to relate the accretion processes in 
neutron star systems and black hole systems, are rare (\S~\ref{Introduction}).
Besides, \citet{Gladstoneetal2007} proposed to connect millisecond pulsars
with verticals, and tried to explain the transition using magnetic field,
because the other two neutron star LMXBs with large scale hysteresis (known
at that time) are verticals (see \S~\ref{Introduction}). These authors 
speculated that millisecond pulsars would be observed to be verticals, if they
reach a high enough luminosity to make a transition. They also suggested that,
among millisecond pulsars, verticals and diagonals (\S~\ref{Introduction}), the first group has 
the highest magnetic field, and verticals have a small magnetic field to affect
the accretion flow indirectly. The properties of IGR J17480--2446 argue against these
suggestions. This is because it is a pulsar (although not a millisecond pulsar)
and a diagonal (see Fig.~\ref{CD}), which has shown large scale hysteresis.
Furthermore, since IGR J17480--2446 is a pulsar, it should have a sufficiently
high magnetic field, even though it is a diagonal.

Fig.~\ref{schematic} shows that, while the peak luminosities of the soft state of 
Aquila X-1, EXO 1745--248 and IGR J17480--2446 are very similar, their 
hard-to-soft transition luminosities are very different. This somewhat argues 
against the correlation between these two kinds of luminosities mentioned in
\citet{YuYan2009} (but note that we have considered only three outbursts). More
importantly, Fig.~\ref{schematic} suggests that the spectral evolution throughout
an outburst does not, at least entirely, depend on the peak luminosity of the outburst.

\begin{figure*}
\centering
\includegraphics*[width=9cm]{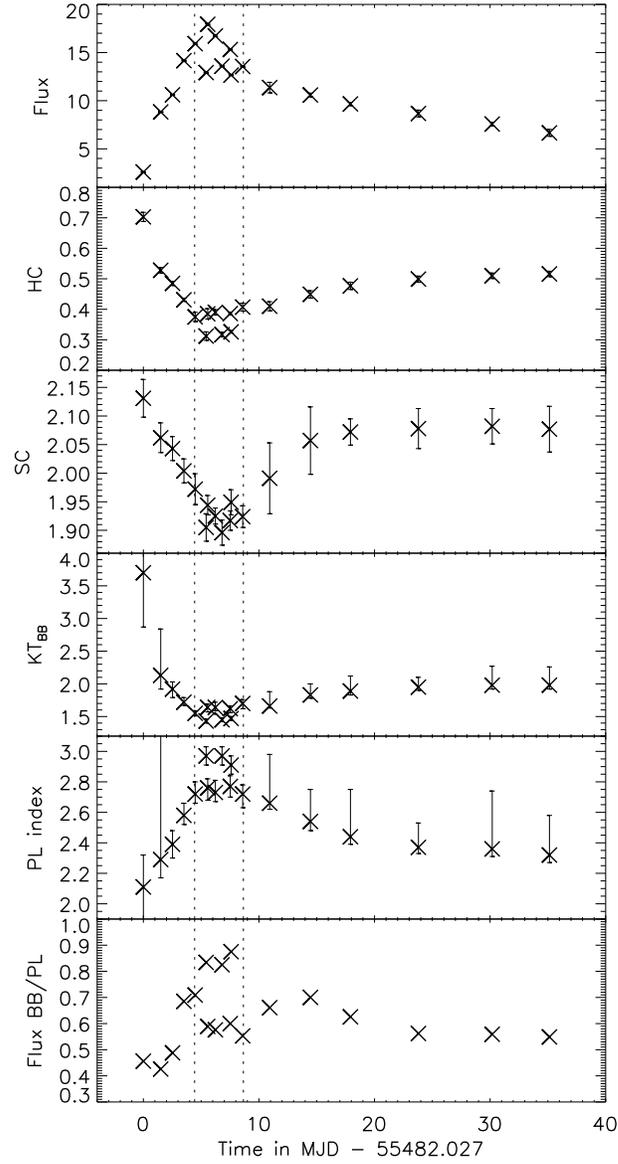}
\caption{Variations of unabsorbed non-burst (persistent) flux (in $10^{-9}$ ergs cm$^{-2}$ s$^{-1}$
in $3-15$ keV), hard colour, soft colour, temperature (keV) of the blackbody spectral
component, index of the powerlaw spectral component, and the ratio of unabsorbed
blackbody flux to unabsorbed powerlaw flux (in $3-15$ keV) with time (day)
during the 2010 outburst of IGR J17480--2446 (\S~\ref{DataAnalysisandResults} 
and \ref{Discussion}). $1\sigma$ error bars for hard colour and soft colour,
and 90\% errors for other parameters are shown. Each point is for one segment
(Table~\ref{tableburst}). This figure shows clear
correlations among all these parameters. The two dotted vertical lines show the 
time range of the most intense segment 5 (Table~\ref{tableburst}).
In this time range, there are two branches roughly for each parameter because of
the systematic difference between the harder and softer parts of the `Z' tracks.
\label{spectime}}
\end{figure*}

\begin{figure*}
\centering
\includegraphics*[width=9cm]{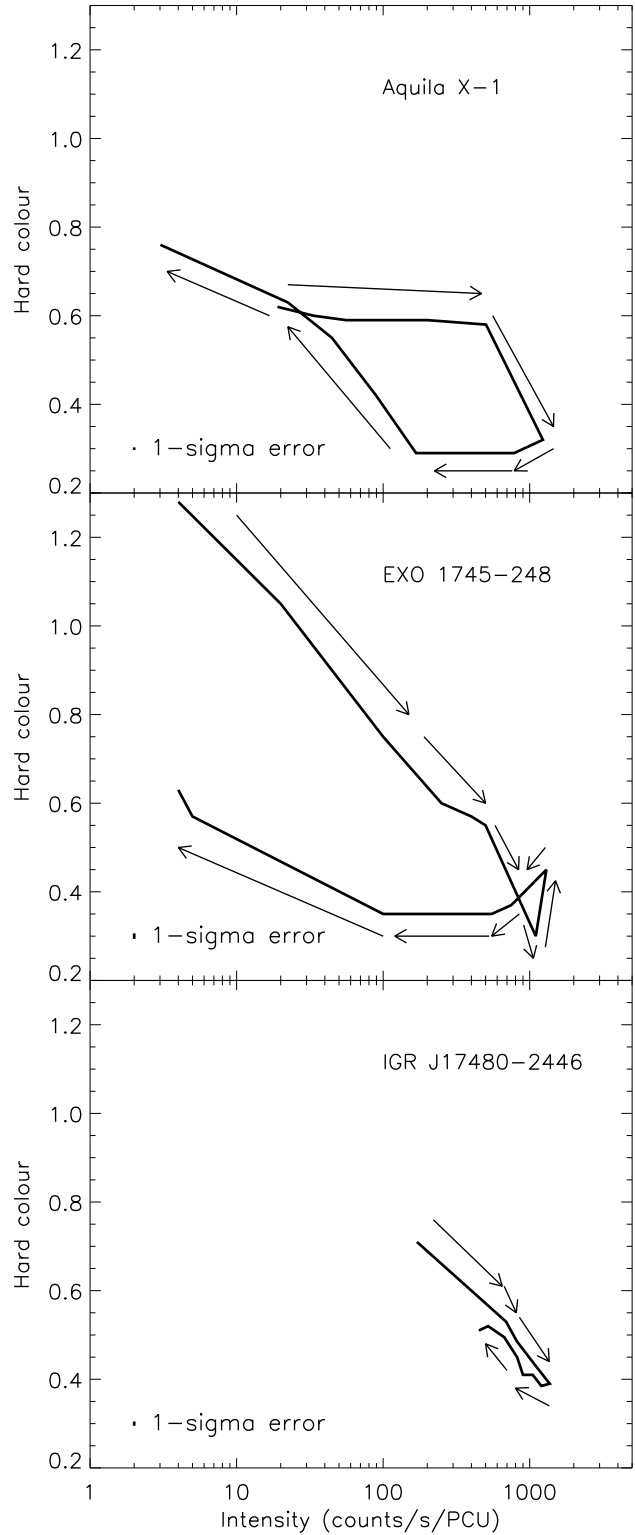}
\caption{Schematic representations of HID diagrams of Aquila X-1 (data during September 24 -- 
November 30 2000), EXO 1745--248 \citep{MukherjeeBhattacharyya2011} and IGR J17480--2446
(using Fig.~\ref{HID}). 
Typical 1$\sigma$ errors in hard colours are shown. Note that both EXO 1745--248 and 
IGR J17480--2446 are in the globular cluster Terzan 5, and hence their distances are about
5.5 kpc \citep{Ortolanietal2007}. On the other hand, since the distance of Aquila X-1 is
about 5.2 kpc \citep{JonkerNelemans2004}, we have normalized its intensity for a distance of 5.5 kpc for a proper comparison with the other two sources (see \S~\ref{Discussion}).
\label{schematic}}
\end{figure*}

We tried to fit the source spectrum with the simplest and the most common single component
models: blackbody, disk-blackbody and powerlaw, none of which worked. Moreover, the continuum
spectra could not be described with a double thermal model, or a two-component 
thermal+non-thermal model involving
a disk-blackbody. The thermal+non-thermal model blackbody+powerlaw was the simplest
model which described the continuum spectra well throughout the outburst. Such a model is not
unusual for pulsars. For example, \citet{PoutanenGierlinski2003} described the 
continuum spectrum of the LMXB pulsar SAX J1808.4--3658 with a blackbody and a Comptonization
components. While the former could originate from the neutron star surface,
the latter, of which the powerlaw might be a phenomenological description,
could originate from a radiative shock from close to the stellar surface.
The best-fit blackbody normalization, which is a measure of the blackbody emission
area, roughly increased when the source became softer, and decreased when the
source became harder (Table~\ref{tablespec2}). This could give a clue about
how the emission area at the magnetic pole changed as the source state evolved,
and hence could be useful to probe the neutron star magnetic field strength
and structure, and the interaction of this field with the accretion flow.
However, we note that the blackbody normalization might also change due to
a systematic change in the colour factor, which includes the effects of
scattering and absorption \citep{Londonetal1984}.
The Fig.~\ref{spectime} shows that the evolutions of hard colour and soft colour are correlated with
the persistent flux evolution, which can also be seen from Figs.~\ref{HID} and \ref{CD}.
This figure also shows, that the blackbody temperature and the powerlaw index 
are correlated, which is expected if these two components originated from almost
the same region, for example the magnetic pole. Moreover, these components
helped each other to change the source states. For example, blackbody temperature
decreased and the powerlaw index increased to make the source softer (Fig.~\ref{spectime}).
The strength of the thermal component relative to that of the non-thermal component
increased, as the source became softer and more luminous (Fig.~\ref{spectime}). This
figure also displays how the various spectral parameters smoothly changed as IGR J17480--2446
transitioned between atoll state and `Z' state.

Finally, Table~\ref{tableburst} shows that the burst peak flux and the burst
interval are correlated with the source states (defined with intensity, hard colour and soft colour).
Both peak flux and interval decreased as the source became softer and more intense,
and increased as the source became harder and less intense. The properties of the
bursts were significantly different between the harder (5b1, 5c1, 5d1) and 
the softer (5a, 5b2, 5c2, 5d2) parts of `Z' tracks: while in the harder parts the 
bursts were visible as small blips in the light curves, 
in the softer parts they were not visible. This behaviour 
is interesting, because such disappearance of bursts in the softer parts is not
due to source intensity increase (average intensity is lower for softer parts
of Z-tracks; Table~\ref{tableburst}), but plausibly due to the change of colour values.

\begin{figure*}
\centering
\includegraphics*[width=9cm]{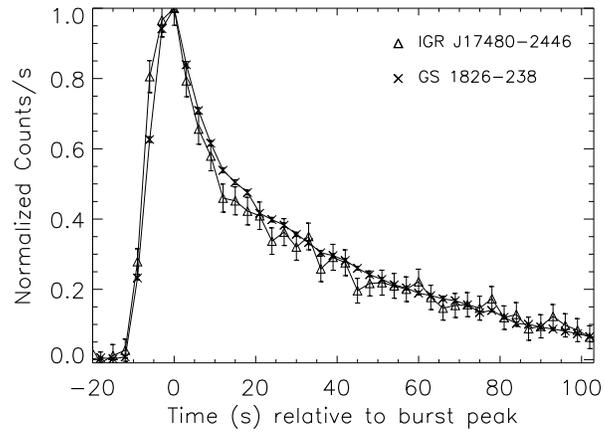}
\caption{Comparison of a typical burst of GS 1826--238 (ObsID: 30054-04-02-01; 1998 June 07 05:31:28) with the October 13 burst of IGR J17480--2446. Each burst light curve is pre-burst
level subtracted and normalized with the peak count rate (see \S~\ref{Discussion}).
\label{shape}}
\end{figure*}

\begin{figure*}
\centering
\includegraphics*[width=9cm]{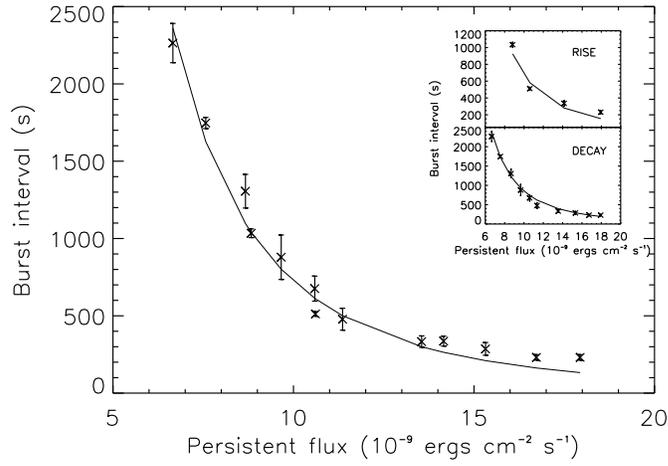}
\caption{Burst interval (with 1$\sigma$ errors) versus non-burst or persistent flux (in $3-15$ keV)
for the 2010 outburst of IGR J17480--2446. In the inset,
similar plots are given for the rising part and the decay part of the outburst. For each plot,
the best-fit powerlaw model has been shown (see \S~\ref{Discussion}).
\label{intervalflux}}
\end{figure*}

Now we ask the question why the burst peak flux, fluence and interval behave with 
the non-burst or persistent emission in the way shown in Table~\ref{tableburst}.
It is theoretically known that at higher accretion rate per unit area (which may
imply higher non-burst emission), the
thermonuclear burning on neutron stars becomes more stable, and the burst interval decreases 
(\citet{StrohmayerBildsten2006} and references therein; \citet{NarayanHeyl2003}). 
That the thermonuclear burning approached the stability as the IGR J17480--2446
intensity increased is supported by the facts that (1) burst intervals smoothly transitioned 
into mHz QPO time periods (Table~\ref{tableburst}; see also \citet{Linaresetal2010}), and
(2) the most popular model of mHz QPO involves the marginally stable nuclear burning
(\S~\ref{Introduction}). Therefore the observed burst behaviour may be qualitatively
expected, and one needs to explain this behaviour quantitatively to find out
why it is not common among bursting LMXBs. Such a theoretical analysis is out of the scope
of this paper, and instead we have compared two properties of IGR J17480--2446
with those of the only other clock-like burster GS 1826--238. A comparison 
between a typical burst light curve of GS 1826--238 with the light curve of the October 13 burst
of IGR J17480--2446 shows that they are very similar (Fig.~\ref{shape}).
This suggests that IGR J17480--2446 is a somewhat GS 1826--238 analogue. 
We have also compared the burst interval versus the persistent flux
plots for IGR J17480--2446 and GS 1826--238 (Fig. 2 of \citet{Thompsonetal2008}). 
We have fitted the plots with a model $\Delta t \propto F_X^{-\beta}$
($\Delta t$ is burst interval and $F_X$ is persistent flux).
From the fitting of the plots 
(entire outburst, rising part and decay part; Fig.~\ref{intervalflux})
for IGR J17480--2446, we found the best-fit $\beta = 2.91\pm0.07, 2.53\pm0.13$
and $2.52\pm0.06$ respectively (but acceptable fit, in terms of reduced $\chi^2$,
has been found only for the data of the decay portion).
On the other hand, \citet{Thompsonetal2008} fitted the plot with the same model, 
and found a best-fit $\beta = 1.05$.
Therefore, for IGR J17480--2446 the burst interval varies more steeply with 
the persistent flux than
for GS 1826--238. However, we note that in order to get $\beta = 1.05$, 
not only \citet{Thompsonetal2008} had to exclude
2003 April data, but also many other data points were significantly away from their
best-fit model curve. For example, only the 1998 June and 2003 April data give
$\beta = 2.65\pm0.95$, which is similar to the value we found for IGR J17480--2446. 
Comparing the Fig. 2 of \citet{Thompsonetal2008} and the Fig.~\ref{intervalflux},
we find that the most striking difference between the bursts
from these two sources is the intervals of IGR J17480--2446 bursts are much smaller
than those of GS 1826--238 bursts. This might be because the IGR J17480--2446 thermonuclear
burning is plausibly closer to the stability, indicated by the observations of mHz QPOs.
Such near-stability accretion rate per unit area may be possible, because the accreted
material is expected to fall on a smaller neutron star surface area for IGR J17480--2446,
which is a pulsar.

\section*{Acknowledgments}

We thank an anonymous referee for constructive comments.

\clearpage
\begin{table*}
 \centering
\caption{Properties of bursts with various spectral states during the 2010 outburst
of IGR J17480--2446 ($1\sigma$ errors are given, unless otherwise mentioned; \S~\ref{DataAnalysisandResults}).
\label{tableburst}}
\begin{tabular}{cccccccccc}
 Seg$^{1}$ & Start & Stop & Intensity$^{3}$ & HC$^{4}$ & SC$^{5}$ & Burst peak & Burst & Burst & Burst\\
    & Time$^{2}$ & Time$^{2}$ &  &  &  & flux$^{6}$ & interval$^{7}$ & duration$^{8}$ & fluence$^{9}$\\
\hline
1 & 10-13T00:12:32 & 10-13T01:05:04 & 185.3$\pm$2.6 & 0.703$\pm$0.015 & 2.131$\pm$0.033 & 9.89$_{-0.78}^{+0.73}$ & - & 120 & 26.70$_{-0.54}^{+0.52}$\\
2 & 10-14T04:28:16 & 10-14T21:05:36 & 690.3$\pm$34.6 & 0.528$\pm$0.009 & 2.062$\pm$0.026 & 5.36$_{-0.56}^{+0.48}$ & 1034$\pm$28 & 105 & 15.33$_{-0.67}^{+0.61}$\\
3 & 10-15T10:24:32 & 10-15T16:00:32 & 832.2$\pm$26.8 & 0.485$\pm$0.007 & 2.043$\pm$0.021 & 2.73$_{-0.23}^{+0.02}$ & 512$\pm$15 & 72 &  6.60$_{-0.29}^{+0.22}$\\
4 & 10-16T11:32:32 & 10-16T14:25:04 & 1110.2$\pm$21.8 & 0.431$\pm$0.005 & 2.004$\pm$0.021 & 2.46$_{-0.27}^{+0.22}$ & 336$\pm$32& 58 &  5.53$_{-0.29}^{+0.24}$\\
5a & 10-17T11:02:24 & 10-17T13:31:28 & 1263.9$\pm$114.4 & 0.375$\pm$0.015 & 1.972$\pm$0.027 & - & - & - & -\\
5b1 & 10-18T07:17:36 & 10-18T10:03:44 & 1410.3$\pm$45.5 & 0.385$\pm$0.017 & 1.944$\pm$0.017 & - & 230.9$\pm$17.6 & - & -\\
,, & 10-18T12:07:28 & 10-18T16:37:36 & ,,         & ,,         & ,,        & - & ,, & - & -\\
5b2 & 10-18T10:32:32 & 10-18T11:45:04 & 1038.2$\pm$142.2 & 0.312$\pm$0.014 & 1.905$\pm$0.024 & - & - & - & -\\
5c1 & 10-19T03:32:32 & 10-19T08:56:32 & 1344.6$\pm$36.0 & 0.390$\pm$0.009 & 1.925$\pm$0.014 & - & 230.9$\pm$17.6 & - & -\\
5c2 & 10-19T19:12:16 & 10-19T20:52:32 & 1088.7$\pm$57.6 & 0.317$\pm$0.008 & 1.896$\pm$0.022 & - & - & - & -\\
5d1 & 10-20T12:41:36 & 10-20T13:41:09 & 1201.7$\pm$17.7 & 0.386$\pm$0.004 & 1.917$\pm$0.017 & - & 285.7$\pm$40.8 & - & -\\
5d2 & 10-20T14:15:28 & 10-20T15:15:09 & 1013.7$\pm$65.8 & 0.326$\pm$0.004 & 1.949$\pm$0.022 & - & - & - & -\\
5e & 10-21T13:42:24 & 10-21T16:21:36 & 1053.0$\pm$20.4 & 0.407$\pm$0.013 & 1.924$\pm$0.019 & - & 333.3$\pm$36.6 & - & -\\
6 & 10-22T13:23:12 & 10-25T08:49:36 & 898.2$\pm$34.7 & 0.410$\pm$0.015 & 1.991$\pm$0.062 & 2.39$_{-0.19}^{+0.14}$ & 478$\pm$71 & - &  4.04$_{-0.15}^{+0.12}$\\
7 & 10-26T08:07:28 & 10-28T14:58:40 & 814.1$\pm$24.2 & 0.449$\pm$0.012 & 2.057$\pm$0.059 & 3.26$_{-0.20}^{+0.17}$ & 676$\pm$81 & 78$\pm$8 &  9.44$_{-0.24}^{+0.21}$\\
8 & 10-29T12:58:24 & 11-01T08:33:36 & 731.9$\pm$17.0 & 0.477$\pm$0.012 & 2.072$\pm$0.023 & 3.67$_{-0.19}^{+0.16}$ & 879$\pm$144 & 74$\pm$23 &  9.39$_{-0.21}^{+0.19}$\\
9 & 11-02T09:29:20 & 11-09T06:21:36 & 652.7$\pm$28.6 & 0.499$\pm$0.010 & 2.078$\pm$0.035 & 5.23$_{-0.18}^{+0.16}$ & 1306$\pm$109 & 114$\pm$11 & 16.48$_{-0.21}^{+0.19}$\\
10 & 11-10T03:26:24 & 11-14T06:32:32 & 569.4 $\pm$17.4 & 0.510$\pm$0.009 & 2.082$\pm$0.031 & 7.44$_{-0.25}^{+0.23}$ & 1747$\pm$36 & 126$\pm$7 & 25.56$_{-0.36}^{+0.32}$\\
11 & 11-15T03:26:24 & 11-19T04:24:32 & 494.0 $\pm$28.0 & 0.516$\pm$0.009 & 2.077$\pm$0.040 & 8.39$_{-0.36}^{+0.34}$ & 2264$\pm$127 & 132$\pm$7 & 30.67$_{-0.89}^{+0.76}$\\
\hline
\end{tabular}
\begin{flushleft}
$^1$Spectral segment numbers (\S~\ref{DataAnalysisandResults}).\\
$^2$Start and stop calendar times of segments (year is 2010).\\
$^3$Mean non-burst count rate in a segment (see \S~\ref{DataAnalysisandResults} for definition).\\
$^4$Mean hard colour in a segment (see \S~\ref{DataAnalysisandResults} for definition).\\
$^5$Mean soft colour in a segment (see \S~\ref{DataAnalysisandResults} for definition).\\
$^6$Mean unabsorbed burst peak flux (in $10^{-9}$ ergs cm$^{-2}$ s$^{-1}$) 
within $3-15$ keV in a segment with 90\% errors. 
Bursts could not be clearly identified in segment 5 (\S~\ref{DataAnalysisandResults}).\\
$^7$ Mean burst interval (in s) in a segment. There is only one burst in segment 1.
For segment 5, some intervals have been estimated from power spectra (\S~\ref{Bursts}).
For others in segment 5, the peaks in power spectra corresponding to bursts could not be cleanly identified.\\
$^8$Mean duration (in s) of bursts in a segment (from the time corresponding to ∼5\% of peak count rate during rise to the time corresponding to ∼5\% of peak count rate
during decay). For first four segments, there are not enough number of bursts to estimate
standard deviations. For the 6th segment, the duration could not be reliably estimated due to intensity fluctuations.\\
$^9$Mean total energy (in $10^{-8}$ ergs cm$^{-2}$) within $3-15$ keV of bursts in a segment
(errors are 90\%).\\

\end{flushleft}
\label{Properties}
\end{table*}

\clearpage
\begin{table*}
 \centering
\caption{Best-fit spectral parameters, unabsorbed total flux (in $3-15$ keV) and the ratio of unabsorbed
blackbody flux to unabsorbed powerlaw flux (in $3-15$ keV) from the spectral fitting of the data of 
each day of each segment (Table~\ref{tableburst}) with the XSPEC model 
{\tt phabs*(bbodyrad+powerlaw+Gaussian)} (\S~\ref{DataAnalysisandResults}).
\label{tablespec2}}
\begin{tabular}{ccccccccc}
Seg$^{1}$ & Date$^{2}$ & Start & Stop & Blackbody & Blackbody      & Powerlaw & Unabsorved flux & FluxBB/FluxPL\\
    & & time$^{2}$ & time$^{2}$ &temp (keV) & normalization$^{3}$ & index$^{4}$  & (10$^{-9}$ ergs s$^{-1}$ cm$^{-2}$) & \\
\hline
1 & 2010-10-13 & 00:12:32 & 01:05:04 & 3.70$_{-0.83}^{+0.38}$ &   0.64$_{- 0.33}^{+ 0.27}$ & 2.11$_{-0.36}^{+0.21}$ &  2.58$_{- 0.02}^{+ 0.01}$ & 0.456\\
2 & 2010-10-14 & 04:28:16 & 05:23:44 & 2.13$_{-0.21}^{+0.71}$ &  13.80$_{- 7.10}^{+ 7.53}$ & 2.29$_{-0.12}^{+0.92}$ &  8.83$_{- 0.04}^{+ 0.05}$ & 0.426\\
3 & 2010-10-15 & 10:24:32 & 11:06:40 & 1.92$_{-0.13}^{+0.11}$ &  27.52$_{- 2.77}^{+11.23}$ & 2.39$_{-0.09}^{+0.09}$ & 10.61$_{- 0.05}^{+ 0.05}$ & 0.488\\
4 & 2010-10-16 & 11:32:32 & 12:24:32 & 1.72$_{-0.05}^{+0.07}$ &  71.48$_{-11.86}^{+ 9.67}$ & 2.58$_{-0.06}^{+0.08}$ & 14.16$_{- 0.06}^{+ 0.06}$ & 0.685\\
5a & 2010-10-17 & 11:02:24 & 12:14:56 & 1.55$_{-0.04}^{+0.04}$ & 128.10$_{-20.26}^{+22.18}$ & 2.72$_{-0.06}^{+0.08}$ & 15.92$_{- 0.07}^{+ 0.08}$ & 0.709\\
5b1 & 2010-10-18 & 13:33:36 & 16:37:36 & 1.64$_{-0.07}^{+0.05}$ & 100.11$_{-14.92}^{+28.16}$ & 2.76$_{-0.08}^{+0.06}$ & 17.94$_{- 0.09}^{+ 0.08}$ & 0.589\\
5b2 & 2010-10-18 & 10:32:32 & 11:45:04 & 1.43$_{-0.03}^{+0.03}$ & 164.09$_{-19.15}^{+23.72}$ & 2.97$_{-0.06}^{+0.06}$ & 12.92$_{- 0.06}^{+ 0.06}$ & 0.834\\
5c1 & 2010-10-19 & 05:10:24 & 06:14:08 & 1.64$_{-0.05}^{+0.08}$ &  93.91$_{-21.17}^{+16.62}$ & 2.73$_{-0.06}^{+0.08}$ & 16.73$_{- 0.07}^{+ 0.08}$ & 0.576\\
5c2 & 2010-10-19 & 19:12:16 & 20:52:32 & 1.45$_{-0.03}^{+0.03}$ & 159.01$_{-18.84}^{+21.54}$ & 2.97$_{-0.06}^{+0.06}$ & 13.59$_{- 0.06}^{+ 0.07}$ & 0.825\\
5d1 & 2010-10-20 & 12:41:36 & 13:41:09 & 1.62$_{-0.06}^{+0.04}$ &  91.50$_{-10.96}^{+19.93}$ & 2.77$_{-0.07}^{+0.07}$ & 15.32$_{- 0.07}^{+ 0.07}$ & 0.600\\
5d2 & 2010-10-20 & 14:15:28 & 15:15:09 & 1.47$_{-0.03}^{+0.03}$ & 142.41$_{-16.62}^{+20.60}$ & 2.91$_{-0.06}^{+0.06}$ & 12.67$_{- 0.06}^{+ 0.06}$ & 0.875\\
5e & 2010-10-21 & 13:42:24 & 16:21:36 & 1.70$_{-0.08}^{+0.06}$ &  62.41$_{- 9.97}^{+19.36}$ & 2.72$_{-0.09}^{+0.06}$ & 13.55$_{- 0.06}^{+ 0.06}$ & 0.553\\
6 & 2010-10-22 & 13:23:12 & 14:43:28 & 1.58$_{-0.05}^{+0.39}$ &  85.79$_{-57.37}^{+15.03}$ & 2.70$_{-0.07}^{+0.26}$ & 12.15$_{- 0.06}^{+ 0.06}$ & 0.653\\
6 & 2010-10-23 & 03:20:16 & 04:14:08 & 1.72$_{-0.07}^{+0.60}$ &  47.23$_{-35.19}^{+11.13}$ & 2.63$_{-0.07}^{+0.97}$ & 11.08$_{- 0.05}^{+ 0.06}$ & 0.517\\
6 & 2010-10-24 & 07:32:32 & 08:45:04 & 1.66$_{-0.06}^{+0.05}$ &  66.04$_{- 9.95}^{+14.11}$ & 2.66$_{-0.07}^{+0.07}$ & 11.27$_{- 0.05}^{+ 0.05}$ & 0.686\\
6 & 2010-10-25 & 05:50:24 & 06:34:40 & 1.69$_{-0.06}^{+0.53}$ &  63.85$_{-44.28}^{+14.89}$ & 2.63$_{-0.09}^{+0.77}$ & 10.92$_{- 0.05}^{+ 0.05}$ & 0.787\\
7 & 2010-10-26 & 08:07:28 & 09:15:28 & 1.76$_{-0.07}^{+0.49}$ &  42.34$_{-30.58}^{+ 8.83}$ & 2.56$_{-0.07}^{+0.62}$ & 10.56$_{- 0.05}^{+ 0.06}$ & 0.552\\
7 & 2010-10-27 & 09:12:16 & 12:26:40 & 1.87$_{-0.10}^{+0.06}$ &  41.69$_{- 5.57}^{+11.68}$ & 2.58$_{-0.12}^{+0.09}$ & 10.80$_{- 0.05}^{+ 0.04}$ & 0.804\\
7 & 2010-10-28 & 12:16:16 & 14:58:40 & 1.85$_{-0.10}^{+0.07}$ &  40.33$_{- 6.19}^{+11.96}$ & 2.49$_{-0.11}^{+0.09}$ & 10.42$_{- 0.05}^{+ 0.04}$ & 0.745\\
8 & 2010-10-29 & 12:58:24 & 14:27:28 & 1.83$_{-0.08}^{+0.36}$ &  39.03$_{-26.55}^{+ 5.07}$ & 2.47$_{-0.09}^{+0.78}$ &  9.87$_{- 0.04}^{+ 0.05}$ & 0.697\\
8 & 2010-10-30 & 08:21:20 & 11:00:32 & 1.86$_{-0.09}^{+0.84}$ &  35.02$_{-23.93}^{+ 9.09}$ & 2.45$_{-0.09}^{+0.94}$ &  9.70$_{- 0.04}^{+ 0.05}$ & 0.669\\
8 & 2010-10-31 & 07:52:16 & 10:35:28 & 1.93$_{-0.12}^{+0.09}$ &  27.10$_{- 4.75}^{+ 9.98}$ & 2.43$_{-0.11}^{+0.08}$ &  9.57$_{- 0.04}^{+ 0.04}$ & 0.596\\
8 & 2010-11-01 & 05:50:24 & 08:33:36 & 1.95$_{-0.14}^{+0.10}$ &  24.14$_{- 4.54}^{+10.24}$ & 2.41$_{-0.11}^{+0.08}$ &  9.48$_{- 0.04}^{+ 0.04}$ & 0.541\\
9 & 2010-11-02 & 09:29:20 & 12:13:36 & 1.92$_{-0.12}^{+0.55}$ &  23.68$_{-14.31}^{+ 8.67}$ & 2.43$_{-0.09}^{+0.68}$ &  8.79$_{- 0.04}^{+ 0.04}$ & 0.520\\
9 & 2010-11-03 & 04:22:24 & 06:03:28 & 1.89$_{-0.08}^{+0.29}$ &  27.76$_{-12.08}^{+ 6.01}$ & 2.37$_{-0.08}^{+0.21}$ &  8.99$_{- 0.04}^{+ 0.04}$ & 0.580\\
9 & 2010-11-04 & 03:50:24 & 07:02:40 & 1.97$_{-0.14}^{+0.10}$ &  22.27$_{- 4.12}^{+ 9.14}$ & 2.39$_{-0.11}^{+0.08}$ &  8.94$_{- 0.04}^{+ 0.04}$ & 0.551\\
9 & 2010-11-05 & 05:22:24 & 08:06:40 & 2.00$_{-0.15}^{+0.10}$ &  21.98$_{- 3.90}^{+ 9.12}$ & 2.40$_{-0.12}^{+0.09}$ &  8.91$_{- 0.04}^{+ 0.04}$ & 0.581\\
9 & 2010-11-06 & 06:05:20 & 09:10:40 & 1.93$_{-0.10}^{+0.63}$ &  24.94$_{-15.05}^{+ 6.70}$ & 2.35$_{-0.09}^{+0.67}$ &  8.68$_{- 0.04}^{+ 0.04}$ & 0.597\\
9 & 2010-11-08 & 03:39:12 & 06:47:28 & 1.93$_{-0.09}^{+0.55}$ &  23.85$_{-15.44}^{+ 6.18}$ & 2.32$_{-0.08}^{+0.54}$ &  8.33$_{- 0.03}^{+ 0.04}$ & 0.578\\
9 & 2010-11-09 & 03:02:24 & 06:21:36 & 2.00$_{-0.16}^{+0.10}$ &  18.43$_{- 3.45}^{+ 8.90}$ & 2.35$_{-0.12}^{+0.08}$ &  8.07$_{- 0.04}^{+ 0.03}$ & 0.526\\
10 & 2010-11-10 & 03:26:24 & 06:21:36 & 1.99$_{-0.15}^{+0.81}$ &  18.53$_{-10.82}^{+ 7.64}$ & 2.41$_{-0.11}^{+1.02}$ &  7.73$_{- 0.03}^{+ 0.04}$ & 0.549\\
10 & 2010-11-11 & 02:07:12 & 04:49:36 & 1.94$_{-0.10}^{+0.59}$ &  20.72$_{-12.04}^{+ 5.62}$ & 2.34$_{-0.09}^{+0.51}$ &  7.65$_{- 0.03}^{+ 0.03}$ & 0.559\\
10 & 2010-11-12 & 03:35:12 & 06:13:36 & 2.03$_{-0.16}^{+0.57}$ &  17.02$_{- 9.70}^{+ 7.16}$ & 2.39$_{-0.12}^{+0.98}$ &  7.62$_{- 0.03}^{+ 0.04}$ & 0.553\\
10 & 2010-11-13 & 04:44:16 & 07:00:32 & 2.01$_{-0.15}^{+0.64}$ &  18.23$_{-11.31}^{+ 7.25}$ & 2.37$_{-0.12}^{+1.08}$ &  7.57$_{- 0.03}^{+ 0.04}$ & 0.583\\
10 & 2010-11-14 & 03:56:16 & 06:32:32 & 1.95$_{-0.09}^{+0.56}$ &  19.26$_{-11.54}^{+ 4.67}$ & 2.30$_{-0.08}^{+0.45}$ &  7.27$_{- 0.03}^{+ 0.03}$ & 0.553\\
11 & 2010-11-15 & 03:26:24 & 06:29:04 & 2.00$_{-0.12}^{+0.75}$ &  16.69$_{-10.32}^{+ 5.40}$ & 2.34$_{-0.10}^{+0.26}$ &  7.09$_{- 0.03}^{+ 0.04}$ & 0.550\\
11 & 2010-11-16 & 23:21:20 & 02:27:28 & 1.91$_{-0.10}^{+0.61}$ &  18.77$_{-14.74}^{+ 5.09}$ & 2.25$_{-0.08}^{+0.84}$ &  6.76$_{- 0.03}^{+ 0.03}$ & 0.527\\
11 & 2010-11-18 & 00:43:12 & 03:03:28 & 1.94$_{-0.08}^{+0.52}$ &  18.07$_{-11.73}^{+ 3.76}$ & 2.28$_{-0.07}^{+0.54}$ &  6.55$_{- 0.03}^{+ 0.03}$ & 0.568\\
11 & 2010-11-19 & 01:31:12 & 04:24:32 & 2.06$_{-0.16}^{+0.11}$ &  13.08$_{- 2.44}^{+ 5.67}$ & 2.39$_{-0.12}^{+0.10}$ &  6.24$_{- 0.03}^{+ 0.03}$ & 0.551\\
\hline
\end{tabular}
\begin{flushleft}
$^1$Spectral segment numbers (\S~\ref{DataAnalysisandResults}).\\
$^2$Date, start and stop calendar times of the data used for the spectral analysis.\\
$^3$$R_{\rm km}^2/D_{10}^2$, where $R_{\rm km}$ is the source radius in km, and $D_{10}$ is the distance to the source in units of 10 kpc.\\
$^4$The powerlaw formula is $KE^{-\alpha}$, where $E$ is in keV,  
$\alpha$ is the powerlaw index, and $K$ is photons/keV/cm$^2$/s at 1 keV.\\
\end{flushleft}
\label{Properties}
\end{table*}

\clearpage

\clearpage

\clearpage

\clearpage

\clearpage

\clearpage

\clearpage

\clearpage

\end{document}